\documentclass[prl,twocolumn,showpacs,superscriptaddress,floatfix]{revtex4}

\usepackage{graphicx}
\usepackage{amsmath}
\usepackage{amssymb}
\usepackage{amsfonts}
\usepackage{amstext}

\usepackage{hyperref}
\usepackage[figure,table]{hypcap} 		
\hypersetup{%
	pdftitle = {Nonequilibrium Dynamics of Anisotropic Large Spins in the Kondo Regime: Time-Dependent Numerical Renormalization Group Analysis.},
	pdfsubject = {},
	pdfauthor = {David Roosen, Maarten Wegewijs, Walter Hofstetter},
	pdfkeywords = {},
	pdfcreator = {},
	pdfproducer = {LaTeX with hyperref},
	colorlinks = true,
		linkcolor = red,		
		anchorcolor = red,		
		citecolor = red,		
		filecolor = red,		
		menucolor = red,		
		pagecolor = red,		
		urlcolor  = red,			
	breaklinks = true,
	bookmarks = false,
	pdfstartview = FitH,			
	pdfhighlight = /I,			
	pdfpagelayout = OneColumn,		
	hypertexnames=true			
}

\begin{document}

\title{
  Nonequilibrium Dynamics of Anisotropic Large Spins in the Kondo Regime: \\
  Time-Dependent Numerical Renormalization Group Analysis  
}
\author{David Roosen}
\affiliation{Institut f\"ur Theoretische Physik, Johann Wolfgang Goethe-Universit\"at, 60438 Frankfurt/Main, Germany}
\author{Maarten R. Wegewijs}
\affiliation{Institut f\"ur Festk{\"o}rper-Forschung -- Theorie 3, Forschungszentrum J{\"u}lich, 52425 J{\"u}lich,  Germany}
\affiliation{Institut f\"ur Theoretische Physik A, RWTH Aachen, 52056 Aachen, Germany}
\author{Walter Hofstetter}
\affiliation{Institut f\"ur Theoretische Physik, Johann Wolfgang Goethe-Universit\"at, 60438 Frankfurt/Main, Germany}

\date{May 25, 2007}

\begin{abstract}
  We investigate the \emph{time-dependent} Kondo effect in a
  single-molecule magnet (SMM) strongly coupled to metallic electrodes. 
  Describing the SMM by a Kondo model with large spin $S >
  1/2$, we analyze the underscreening of the local moment and the
  effect of anisotropy terms on the relaxation dynamics of the 
  magnetization. Underscreening by single-channel Kondo
  processes leads to a logarithmically slow relaxation, while finite
  uniaxial anisotropy causes a saturation of the SMM's
  magnetization. Additional transverse anisotropy terms induce
  quantum spin tunneling and a pseudospin-$1/2$ Kondo effect 
  sensitive to the spin parity.
\end{abstract}


\maketitle

\emph{Introduction.} Crystals of single-molecule magnets (SMMs) are
mesoscopic systems, which exhibit fundamental quantum mechanical phenomena
like quantum tunneling of the magnetization (QTM).  The individual SMMs
responsible for these effects have large spin $S \sim 10$ and show rich
quantum dynamics due to the anisotropic effective potential induced by
ligand-fields and spin-orbit coupling~\cite{Gatteschi}.  More recently,
transport through individual SMMs coupled to metallic electrodes has been
measured for the first time~\cite{Heersche,Jo}, giving access to many-body
physics in these complex quantum impurity systems.  In particular, the
large spin, in combination with magnetic anisotropy,  was predicted to give
rise to a pseudospin-$1/2$ Kondo effect~\cite{Romeike1} in a single SMM.
In combination with applied magnetic fields, this effect can be used for
transport spectroscopy of an \emph{individual}  SMM in a two-terminal
geometry~\cite{Romeike2}.  Beyond the linear response transport
investigated in these works, one can gain fundamental insight by studying
nonequilibrium magnetization dynamics of an SMM coupled to macroscopic
electrodes.  Recent STM experiments have also demonstrated that single
molecules attached to a substrate can be abruptly switched to a new
conformation with different transport properties by a voltage pulse,
resulting, for instance, in a change in the Kondo temperature~\cite{Iancu06}.
Recent advances in the field of  quantum computing using electron spins in
semiconductor quantum dots concern high-frequency control of magnetic and
electric signals. In particular, recently it has been shown
experimentally~\cite{Nowack07} that via oscillating electric fields
spin transitions can be induced locally by spin-orbit coupling. There are
prospects for applying such techniques to three-terminal single-molecule
devices although technical challenges remain.
\par

Motivated by these advances, here we theoretically analyze the
nonequilibrium spin dynamics of a SMM in response to a sudden change of
the magnetic field strength, with a special focus on the manifestation of
Kondo screening in the time domain.   
\begin{figure}[t] 
    \centering 
      { 
      \includegraphics[width=0.75\linewidth]{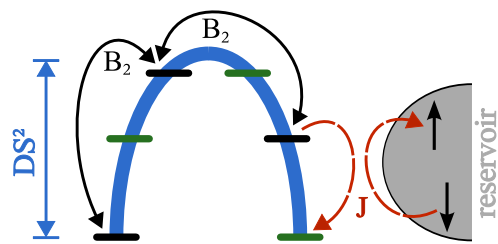} 
      }
      \vspace{-0.4cm}
      \caption[Sketch of SMM with conduction band.] 
      {\label{fig:SMM_sketch}
       Energy diagram of the SMM.  The levels indicate spin eigenstates
       $|S,M\rangle$ which are split in energy by the \emph{uniaxial} ($D$)
       anisotropy.  Two types of spin fluctuations occur: processes due to the
       intrinsic \emph{transverse} anisotropy ($B_2$) , resulting in QTM, and
       exchange processes involving the conduction band electrons ($J$),
       leading to the Kondo effect.  
      }
    \vspace{-0.4cm}
\end{figure}
The accurate real-time calculation of impurity observables like the
magnetization is crucial for developing a general understanding of
time-resolved Kondo spectroscopy of SMMs, for instance using transport
measurements~\cite{Romeike2}.  More generally, proposals for spintronic
operation of SMM-based
devices~\cite{RomeikeSET,Elste,Misiorny,Gonzalez,Lehmann07}  rely on a
detailed understanding of the magnetization dynamics.  Note that here we
consider pulsed magnetic fields as opposed to the experimentally more
easily accessible oscillating fields, since the former are at the moment
more conveniently treated using numerical renormalization group (NRG)
methods \cite{Wilson75,Hofstetter00,Costi,AndersSchiller}. 
In order to analyze the effects of pulsed fields we use the time-dependent
numerical renormalization group method (TD-NRG) recently introduced by
Anders and Schiller~\cite{AndersSchiller}.  The advantage of this approach
is that the time evolution of any quantum impurity system after a sudden
perturbation at $t=0$ can be calculated in a numerically exact way on
arbitrarily long time scales without accumulating any discretization error
that scales with the elapsed time, in contrast, e.g.\, to the adaptive
time-dependent DMRG~\cite{t-DMRG}.  For a detailed description of the
TD-NRG method we refer to~\cite{AndersSchiller}.  For all the results
presented below we have exhaustively checked that the dependence on the
method's parameters is negligible (such as the discretization parameter
$\Lambda$ or the number of states $N_{Level}$ taken into account in each
iteration).  We employ units in which $ g = \hbar = k_B = W = 1$, where $W$
is half the width of the conduction band.
\par

\emph{Model.} We consider a SMM as a (generalized) quantum impurity coupled
to a conduction band of metallic electrodes :
$\mathcal{H} = \mathcal{H}_{cb} +
\mathcal{H}_{int} +
\mathcal{H}_{imp} $ \cite{Romeike1}. 
The conduction band and the interaction term in the Hamiltonian read 
$ \mathcal{H}_{cb} = \sum_{k, \mu} \varepsilon_{k} c^{\dagger}_{k \mu}
c^{\phantom{\dagger}}_{k \mu}, \mathcal{H}_{int} = J~\mathbf{S} \cdot
\mathbf{s} $, 
where $\mathbf{s}$ is the spin of the most localized electronic orbital in
the conduction band which is coupled to the impurity spin $\mathbf{S}$
(e.g.\, $S = 10$ for the single-molecule magnet Mn$_{12}$ and $ S  = 8$ in
case of Fe$_8$).  Below we systematically investigate different models
describing the impurity $\mathcal{H}_{imp}$ (see
Fig.~\ref{fig:SMM_sketch}). The time scale on which physical quantities
vary throughout this work typically ranges from $10^3$ to $10^9$ times
$\hbar/ W$ corresponding to 1 ps -- 1 $\mu$s for typical values of the spin,
exchange interaction and anisotropy parameters.
\par

\begin{figure}[t]
  \centering
    {
    \includegraphics[width=0.96\linewidth]{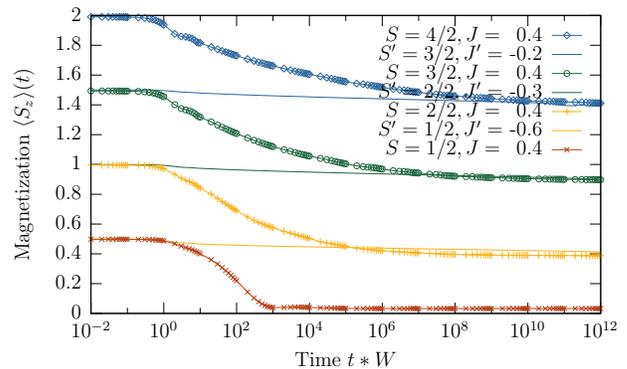}
    }
    \vspace{-0.4cm}
    \caption[Comparison between the relaxation of $S$ and $S'$ with $J'\neq -J$.]
    {\label{fig:mM_under}
    The logarithmic decay of $\langle S_z \rangle (t)$ due to
    underscreening of the local moment for different spin sizes $S$
    ($J=0.4$) is compared to the relaxation of a spin $S' = S - 1/2$, whose
    effective ferromagnetic coupling $J'<0$ is chosen according to
    \protect\cite{Koller}.  (NRG parameters: $\Lambda = 2.5$, $N = 60$,
    $N_{Level} = 2000$).
    }
  \vspace{-0.4cm}
\end{figure}

\emph{Isotropic Kondo model.} Let us first investigate an isotropic spin
$S$ with antiferromagnetic exchange $J>0$ induced by electron tunneling,
i.e.\, $\mathcal{H}_{imp} = 0$. For a single electron with spin $S = 1/2$ at
zero temperature and zero magnetic field the local moment is completely
screened by the Kondo effect, i.e. $\langle S_z \rangle =0$.
Here, however, we consider a large impurity spin $S >1/2$,  as relevant for
SMMs~\cite{Romeike1}, aligned at times $t<0$ by a strong magnetic field
$h_z$ in the longitudinal direction, which is sufficiently strong to
completely polarize the impurity spin.  At $t=0$ the field is switched off
and using the TD-NRG we calculate the screening of the SMM's magnetization
by exchange scattering processes as a function of time, see
Fig.~\ref{fig:mM_under}.  The most prominent observation is that, in
contrast to the $S=1/2$ case, the interaction between the impurity and the
conduction electrons does not quench the magnetic moment on the impurity
completely.  This is referred to as the \textit{underscreened Kondo effect}
\cite{Nozieres80,Cragg79,Bethe_ansatz,Coleman03,Koller}.  In a na\"{\i}ve
picture the equilibrium underscreened Kondo effect can be described as a
two-stage process: first the magnetization is reduced to $S-1/2$ by
screening due to the spin-$1/2$ electrons in the conduction band.  This
effective spin $S'=S-1/2$ then couples ferromagnetically ($J' < 0$) to the
conduction band and becomes asymptotically free at low
energies~\cite{Cragg79}.  Our results in Fig.~\ref{fig:mM_under} show how
this energy dependence of the effective couplings translates into the
real-time dynamics of the impurity spin.  For different values of $S$ and
{$J=0.4$} we observe an initial drop of the magnetization by approximately
$1/2$, followed by a logarithmically slow further relaxation (note that
$\langle S_z \rangle$ is not a conserved quantity and can drop below the
value of $S_z - 1/2$). 
Comparing the low energy excitations for a spin $S$ and antiferromagnetic
coupling to those of a spin $S' = S -1/2$ impurity coupled
ferromagnetically, Koller \emph{et al.}~\cite{Koller} have determined the
value of the renormalized ferromagnetic interaction $J'$ for the
underscreened Kondo effect.  In Fig.~\ref{fig:mM_under} we show that this
correspondence can also be observed in the real-time dynamics, by comparing
the underscreened {$J=0.4$} relaxation dynamics  with a ferromagnetic
$S'=S-1/2$ Kondo impurity and an effective $J'<0$ as calculated
in~\cite{Koller}.  In the ferromagnetic case, we fitted the time-evolution
as $\langle S_z(t) \rangle = A / [\log (t + B) ]$ using $A$ and $B$ as fit
parameters and found very good agreement with our results.  For
antiferromagnetic coupling good asymptotic agreement with the corresponding
ferromagnetic case was found in the time evolution.  This confirms our
identification of time-resolved underscreening.  We observe that the time
scale of the partial screening $S \to S-1/2$ increases with the spin size
$S$. If the exchange interaction $J$ is decreased, the effect is more
dramatic: the partial screening sets in at larger times and its time scale
increases considerably (not shown) in agreement with the non-linear
increase of  the effective coupling $J'$ for small $J$ as found in
\cite{Koller}.
\par

\emph{Kondo model with uniaxial anisotropy.} Single-mo\-lecule magnets are
characterized by additional magnetic anisotropy terms in the impurity part
of the Hamiltonian, as described by the generic model 
$\mathcal{H}_{imp} = -D
S_z^2 - \frac{1}{2} B_2 \left( S_+^{2} + S_-^{2}
\right) - \mathbf{h}\cdot \mathbf{S}$. 
The spin has to overcome an energy barrier of height $DS_z^2$ to reverse its
ground state alignment with the easy axis intrinsic to the SMM.  Ligand
field effects reduce the symmetry group of the molecule about this axis
down to a finite group. The non-commuting terms $\propto B_2$ allowed by
this low symmetry introduce quantum tunneling of the magnetization (QTM)
through the barrier~\cite{Gatteschi}.  The energy levels of an isolated SMM
are sketched in Fig.~\ref{fig:SMM_sketch}.  The resulting model has
recently been studied~\cite{Romeike1,Romeike2} since it is relevant to
transport experiments on SMMs immobilized between metallic
electrodes~\cite{Heersche,Jo}.  
\par

\begin{figure}[t]
  \centering
    {
    \includegraphics[width=0.96\linewidth]{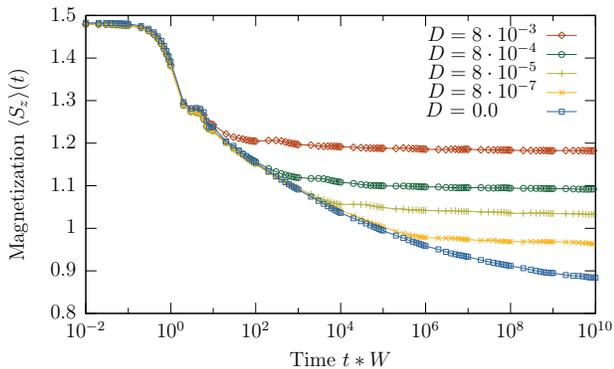}
    }
    \vspace{-0.4cm}
    \caption[Barrier $D$ ``freezes'' the relaxation in the underscreened
    Kondo effect.]
    {\label{fig:mM_D_freeze}
     Suppression of underscreening: the magnetic anisotropy barrier $D>0$
     impedes further relaxation of the SMM's magnetization by the Kondo
     effect ($S=3/2$, $J=0.6$).  (NRG parameters: $\Lambda = 1.8$, $N =
     60$, $N_{Level} = 1000$.)
    }
  \vspace{-0.4cm}
\end{figure}

We first investigate how the development of the underscreened Kondo effect
is hampered by a finite anisotropy barrier $D>0$ in the absence of QTM
($B_2=0$).  When switching off the magnetic field, a clear saturation of
the magnetization on a time scale $\propto 1/D$ is seen in
Fig.~\ref{fig:mM_D_freeze}.  The logarithmic decay characteristic of the
underscreened Kondo effect without anisotropy $D$ is terminated, because
Kondo spin fluctuations with small energy uncertainty are suppressed by the
barrier.  In a renormalization group language, the scaling is cut off at
$D$ and the Kondo screening stays incomplete.
\par

\begin{figure}[t]
  \centering
    {
    \includegraphics[width=0.96\linewidth]{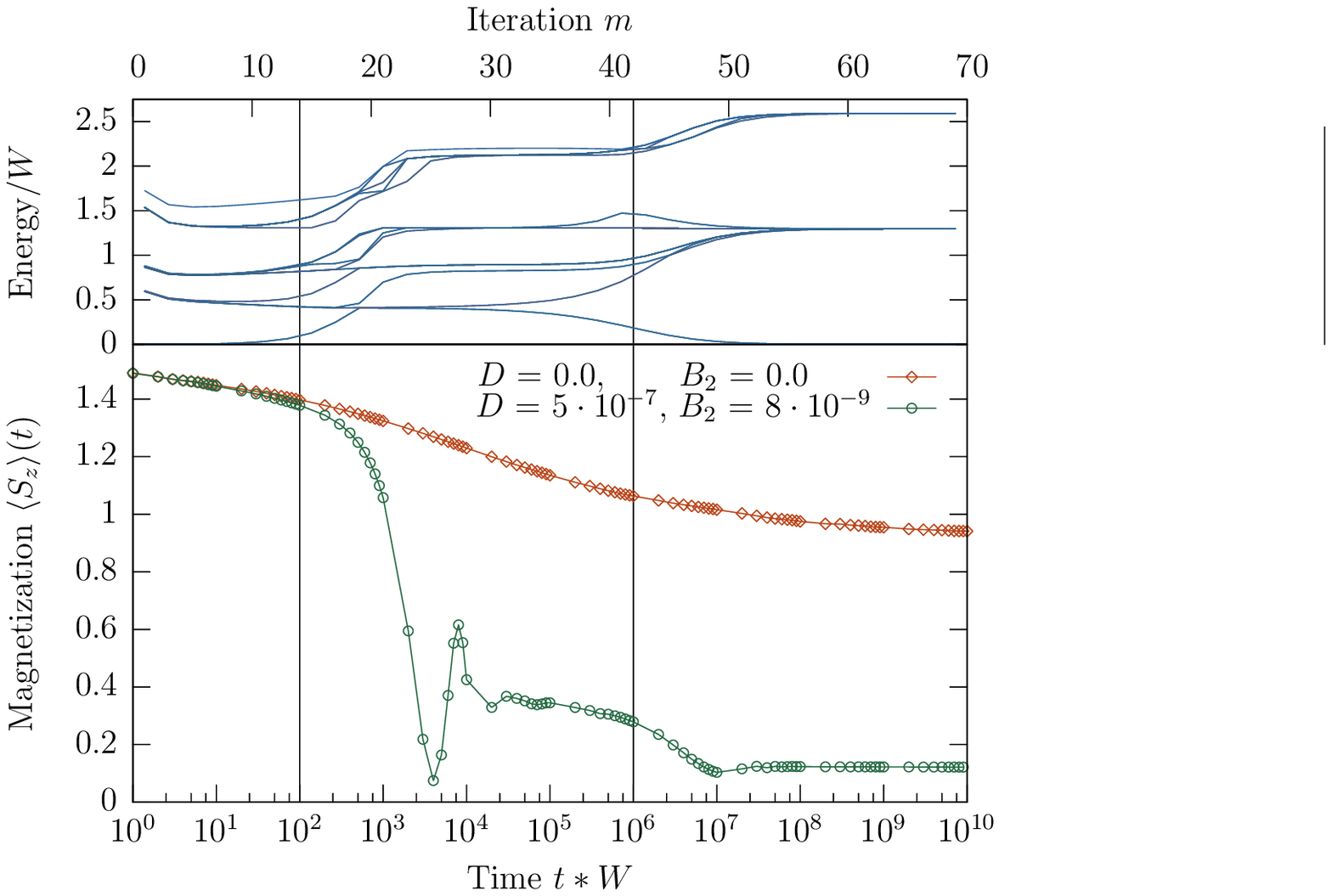}
    }
    \vspace{-0.4cm}
    \caption[time evolution for small B2]
    {\label{fig:mM_small_B2}
     Half-integer spin ($J = 0.2$). Three different regimes in the
     time evolution relate to underscreening ($tW < 100$), buildup of the
     pseudospin-$1/2$ Kondo effect ($100 < tW < 10^{6}$), and approach to the
     strong-coupling fixed point ($tW > 10^{6}$).  Above the time evolution we
     plot the eigenenergies of the NRG Hamiltonian $H_m$ as a function of
     iteration $m$, i.e.\, the spectrum of the many-body system SMM + electrode,
     where conduction-band electrons with energies above $1/\Lambda^m$ have
     been integrated out exactly.  The scales for the two plots are adjusted to
     each other, allowing a direct comparison between the NRG level flow and
     the corresponding real-time evolution.  (NRG parameters: $\Lambda = 2.0$,
     $N = 70$.)
    }
  \vspace{-0.4cm}
\end{figure}

\emph{Kondo model with uniaxial and transverse anisotropy.} The transverse
anisotropy term $B_2$ induces QTM in SMMs in the absence of a magnetic
field. 
We first consider the typical case $B_2 \ll D$ and strong coupling to the
conduction band $J=0.2$ for half-integer spins $S > 1/2$.  The Kondo
effect developing  in this limit  was shown~\cite{Romeike1} to involve all
magnetic excitations of the SMM, since $T_K(S,J,B_2,D)$ exceeds the
magnetic splittings due to the large $J$ value.  As before, we prepare a
spin $S=3/2$ system in an almost completely polarized state $\langle S_z
\rangle (0) \approx 3/2$ at $t<0$,  and monitor the subsequent
time dependence of the magnetization $\langle S_z \rangle (t)$. For very
weak transverse anisotropy ($D = 5 \cdot 10^{-7}$, $B_2 = 8
\cdot10^{-9}$, $J = 0.2$), three different regimes can be distinguished
in the time evolution (cf.\ Fig.~\ref{fig:mM_small_B2}):
(i) Starting at $t \approx 1/W$ the conduction band electrons
  partially screen the spin on the impurity. Both the time evolution
  and the NRG energy levels are the same as for the ``bare''
  underscreened Kondo effect obtained by setting $B_2 = D = 0$.
(ii) The transverse anisotropy $B_2$ introduces a new feature:
  for times $tW > 100$ (i.\ e.\, before the finite barrier $D$
  ``freezes'' the spin) the magnetization starts declining 
  and displays an oscillation around the value $\langle S_z
  \rangle \approx 0.35$.
(iii) The remaining magnetization is quenched by spin
  exchange processes between the conduction band electrons and
  the impurity. The time scale is given by the inverse Kondo temperature of
  the pseudospin-$1/2$ Kondo effect $\tau \propto 1/T_K(S,J,B_2,D)$ in line
  with~\cite{Romeike1,Romeike2}.  The small remnant magnetization at large
  times is due to incomplete thermalization which is unavoidable in a
  finite size system with discretization parameter $\Lambda > 1$; see also
  \cite{AndersSchiller}.
\par

\begin{figure}[t]
  \centering
    {
    \includegraphics[width=0.96\linewidth]{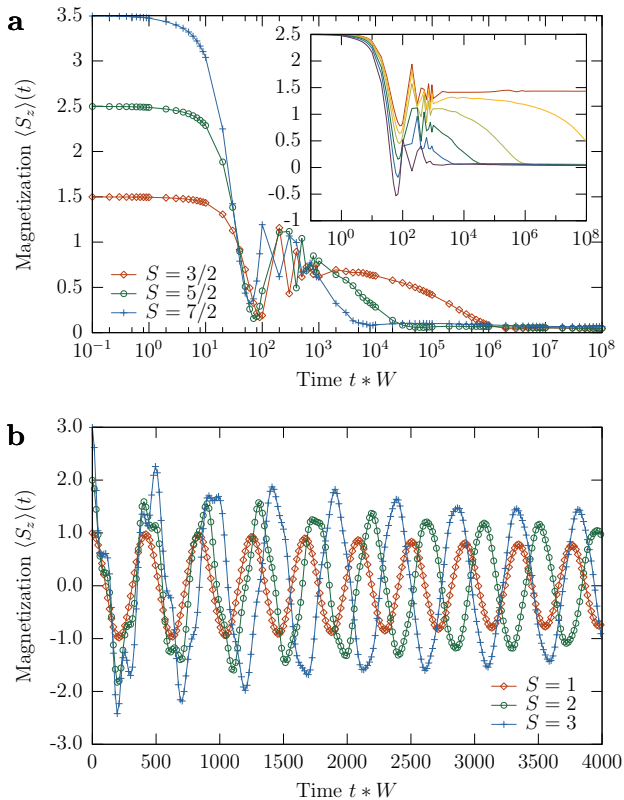}
    }
    \vspace{-0.4cm}
    \caption[Time-dependent magnetization for large $B_2$, weak coupling
      $J$ and different half-integer $S$.]
    {\label{fig:mM_big_B2}
     Magnetization dynamics for weak Kondo coupling $J = 0.1$, large
     QTM, $B_2 =0.75 \cdot D = 7.5 \cdot 10^{-3}$, and 
     different sizes of the spin:
      (a) {\it Half-integer spin: }
	two-stage relaxation, with a time scale for the Kondo effect  	
           strongly dependent on the size of the spin.
	   Inset: dependence on exchange coupling $J$ 
	   [shown for spin $S=5/2$, $J = 0.001$ (red) and $J = 0.05$ (orange) 
           $... 0.15$ (purple) in equidistant steps].
      (b) {\it Integer spin: } 
           the different spin sizes have only little effect on the
           damped Rabi oscillation of the magnetization around zero.
     (NRG parameters: $\Lambda = 2.0$, $N = 65$.)
    }
\end{figure}

In order to study the sensitivity of the observed magnetization dynamics to
the spin size $S$ as well as the exchange coupling $J$, we now consider
large values for the anisotropy barrier $D \sim 10^{-2}W$ and the quantum
tunneling term $B_2 \sim D$, for a weakly coupled SMM ($J = 0.1$).
For {\it half-integer} spin $S$ the quantum tunneling should then reduce
the magnetization to a finite value, before the underscreened Kondo effect
sets in. Our results for the time-dependent magnetization $\langle S_z
\rangle (t)$ indeed display this behavior:
Fig.~\ref{fig:mM_big_B2} (a) shows a clear two-step relaxation and a
strong dependence on the half-integer value of $S$ as well as on the
exchange coupling $J$. At $tW \approx 10$ the quantum tunneling term $B_2$
starts mixing the different magnetic states of the molecule.  The
magnetization is reduced to $\langle S_z \rangle \approx 0.7$ showing
damped Rabi-type oscillations around this value for all spin sizes we
investigated ($S=3/2-7/2$), which are related to the quantum tunneling term
and which disappear for $B_2 = 0$.
For larger times $t$, we again find the pseudospin-$1/2$ Kondo effect to
set the relevant time scale $\tau \propto 1/T_K(S,J,B_2,D)$ for the
complete screening of the molecule's magnetization. Since the Kondo
temperature increases with the size of the spin (cf.~\cite{Romeike1}) the
relaxation is faster for larger spin values $S$ [cf.
Fig.~\ref{fig:mM_big_B2} (a)].
The dependence of the relaxation on the exchange coupling $J$ for fixed
spin $S=5/2$ is shown in the inset of Fig.~\ref{fig:mM_big_B2} (a). For
larger $J$ the oscillations reach negative values.  The narrow plateau
after the first ``drop'' in the magnetization, which is due to the
transverse anisotropy $B_2$, saturates at $\langle S_z \rangle \approx
1.5$, and the time scale for the pseudospin-$1/2$ Kondo effect becomes very
large in the limit of weak coupling $J$.  
\par

For \emph{integer} spin values $S$ the QTM term alone allows the
magnetization to tunnel between the two ground states $|S\rangle$ and
$|\text{-}S\rangle$ of the isolated SMM ($J=0$), resulting in zero net
equilibrium magnetization at low magnetic field.  Therefore, one does not
expect a Kondo effect, because no cooperation of the spin exchange
processes is needed to change the direction of the
magnetization~\cite{Romeike1}.
We indeed observe  a dramatic change in the spin dynamics going from
half-integer spin to the nearest full-integer $S$ for identical parameters
$B_2$, $D$ and $J$ [compare Fig.~\ref{fig:mM_big_B2} (a) and (b)].
$\langle S_z \rangle (t)$ displays an oscillation around zero which dies
out on time scales $\sim 10^{4}/W$ due to damping induced by the
conduction band electrons.
\par

\emph{Conclusions.}
In this work we have analyzed the nonequilibrium magnetization dynamics of
Kondo models with large spin $S > 1/2$ in response to a sudden
perturbation.  For the isotropic Kondo model we found a two-stage
relaxation, which displays the underscreened Kondo effect in the real-time
magnetization dynamics and confirms that the renormalized coupling to the
impurity after a partial screening of the magnetization to $S - 1/2$ is
ferromagnetic $J' < 0$.  For the extended anisotropic Kondo Hamiltonian
describing a single-molecule magnet, we found a suppression of the
underscreened Kondo effect by uniaxial anisotropy.  With additional
transverse anisotropy, relaxation becomes sensitive to the spin
\emph{parity}. For half-integer spin, a complex interplay of tunneling and
spin-screening by conduction electrons leads to a pseudospin-$1/2$ Kondo
effect and a  reduction of the magnetization on long time scales.  In
contrast, for integer spin, the transverse anisotropy only leads to damped
Rabi oscillations without spin screening.
\par

\begin{acknowledgments}
The authors thank Christian Romeike  for discussions.
\end{acknowledgments}

\vspace{-0.4cm}



\begin{references}

\bibitem{Gatteschi}
D.~Gatteschi and R.~Sessoli, Angew.~Chem.~Int.~Ed.~\textbf{42}, 3, 268 (2003) and references therein; W.~Wernsdorfer and R.~Sessoli, Science \textbf{284}, 133 (1999).

\bibitem{Heersche}
H.~Heersche {\it et~al.}, Phys.~Rev.~Lett.~\textbf{96}, 206801 (2006).

\bibitem{Jo}
M.~-H.~Jo {\it et~al.}, Nano Letters \textbf{6}, 9, 2014 (2006).

\bibitem{Romeike1}
C.~Romeike {\it et~al.}, Phys.~Rev.~Lett.~\textbf{96}, 196601 (2006).

\bibitem{Romeike2}
C.~Romeike {\it et~al.}, Phys.~Rev.~Lett.~\textbf{97}, 206601 (2006).

\bibitem{Iancu06}
V.~Iancu, A.~Deshpande, and S.~-W.~Hla, Nano Lett.~\textbf{6}, 820 (2006).

\bibitem{Nowack07}
K.~C.~Nowack {\it et~al.}, Science \textbf{318}, 1430 (2007), and references therein.

\bibitem{RomeikeSET}
C.~Romeike, M.~R.~Wegewijs, and H.~Schoeller, Phys.~Rev.~Lett.~\textbf{96}, 196805 (2006).

\bibitem{Elste}
C.~Timm and F.~Elste, Phys.~Rev.~B \textbf{73}, 235304 (2006).

\bibitem{Misiorny}
M.~Misiorny and J.~Barnas, Europhys.~Lett.~\textbf{78}, 27003 (2007).

\bibitem{Gonzalez}
G.~Gonzalez and M.~N.~Leuenberger, Phys.~Rev.~Lett.~\textbf{98},
256804 (2007).

\bibitem{Lehmann07}
J.~Lehmann and D.~Loss, Phys.~Rev.~Lett.~\textbf{98}, 117203 (2007).

\bibitem{Wilson75}
K.~G.~Wilson, Rev.~Mod.~Phys.~\textbf{47}, 773 (1975).

\bibitem{Hofstetter00}
W.~Hofstetter, Phys.~Rev.~Lett.~\textbf{85}, 1508 (2000).

\bibitem{Costi}
T.~A.~Costi, Phys.~Rev.~B \textbf{55}, 5, 3003 (1997).

\bibitem{AndersSchiller}
F.~Anders and A.~Schiller, Phys.~Rev.~Lett.~\textbf{95}, 196801 (2005), 
and Phys.~Rev.~B \textbf{74}, 245113 (2006).

\bibitem{t-DMRG}
A.~J.~Daley {\it et~al.}, J.~Stat.~Mech.~: Theor.~Exp.~\textbf{P04005} (2004); S.~R.~White and A.~E.~Feiguin, Phys.~Rev.~Lett.~\textbf{93}, 076401 (2004).

\bibitem{Nozieres80}
P.~Nozieres and A.~Blandin, J.~Phys.~(Paris) \textbf{41}, 193 (1980).

\bibitem{Cragg79} 
D.~M.~Cragg and P.~Lloyd, J.~Phys.~C \textbf{12}, L215 (1979).

\bibitem{Bethe_ansatz}
A.~M.~Tsvelick and P.~B.~Wiegmann, Adv.~Phys.~\textbf{32}, 453 (1983);
N.~Andrei, K.~Furuya, and J.~H.~Lowenstein, Rev.~Mod.~Phys.~\textbf{55}, 331 (1983).

\bibitem{Coleman03}
P.~Coleman and C.~Pepin, Phys.~Rev.~B \textbf{68}, 220405 (2003).

\bibitem{Koller}
W.~Koller, A.~C.~Hewson, and D.~Meyer, Phys.~Rev.~B \textbf{72}, 4, 045117, (2005).


\end{references}
\end{document}